# MEG-Derived Functional Tractography, Results for Normal and Concussed Cohorts

Don Krieger, Paul Shepard, Walter Schneider, Sue Beers, Anthony Kontos, Michael Collins, David O. Okonkwo

**Summary**
Measures of neuroelectric activity from each of 18 automatically identified white matter tracts were extracted from resting MEG recordings from a normative (n=588) and a chronic TBI (traumatic brain injury) (n=63) cohort, 60 of whose TBI's were mild. Activity in the TBI cohort was significantly reduced compared with the norms for ten of the tracts, $p < 10^{-6}$ for each. Significantly reduced activity ($p < 10^{-3}$) was seen in more than one tract in seven mTBI individuals and one member of the normative cohort.

**Methods and results**

Imaging studies from the CamCAN lifetime normative dataset [1,2], ages 18-88, and the TEAM-TBI chronically symptomatic cohort [3], ages 21 – 60, were segmented using freesurfer [4] and tracula [5]. For each individual, this yielded xyz coordinates of the volume occupied by each of 18 white matter tracts with 2.4 mm resolution. The tract names are listed in the figure 1 legend.

The referee consensus solver [6] was applied to resting MEG recordings from all subjects and to the 39 6-month follow-up studies obtained for the TEAM-TBI cohort. The average yield of the solver for the CamCAN dataset is 455 neuroelectric currents per 40 msec step through the 560-second resting data stream, each validated at $p < 10^{-12}$, i.e. about 6.3 million validated currents per individual [7]. Each identified current includes its spatial location (1 mm resolution) and its 80 msec vector waveform. The primary output from the solver for the CamCAN dataset is available for download with registration [8].

The neuroelectric currents within the volume of each tract were counted for the entire 560 second recording period. Each count was then normalized to a current density [7] to enable direct comparisons between one individual and another. These values are organized and plotted in Figure 1 to highlight the differences between the cohorts by tract. 10 of the 18 tracts demonstrate reduced population-wide activity in the TBI cohort compared with the CamCAN cohort, $p < 10^{-6}$ for each tract.

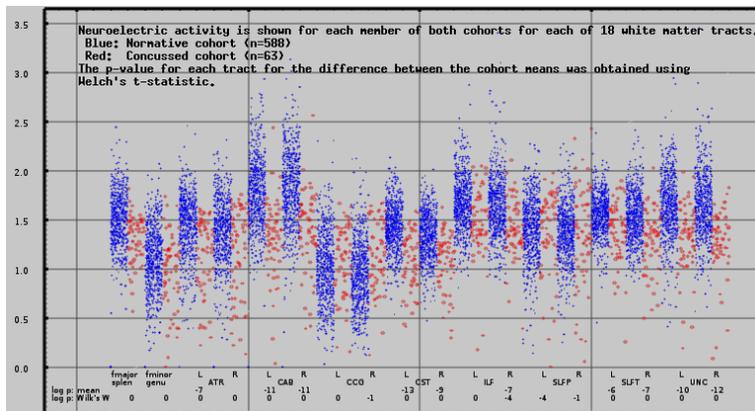

Figure 1. Functional activity in 18 automatically identified tracts is shown in blue for each member of a neurologically normal cohort, ages 18-88, *n* = 588, and in red for each member of a TBI cohort, ages 21 – 60, *n* = 63.

The measures were obtained from resting MEG recordings. The xyz coordinates of the volume occupied by each tract was estimated using tracula (Martinos Imaging Center, Harvard University). Neuroelectric activity density was computed and interpreted as a measure of connectivity strength across the tract.

The $\log_{10}$ p-values for the cohort *means* near the bottom of the figure indicating significant differences ($p < 10^{-6}$) for the indicated tracts. Note that for each significant finding, the mean activity for the TBI cohort is less than that for the normative cohort.

The $\log_{10}$ p-values for the *Wilk's W statistic* near the bottom of the figure indicate those tracts for which the distribution of the CamCAN *means* differs significantly from *normal*, viz. R_ILF and L_SLFP. For these tracts, the confidence that may be placed in the p-values for each individual's tracts shown in Figure 2 is compromised.

The tracts are the forceps major and minor, i.e. corpus callosum splenium and genu, left and right anterior thalamic radiations (ATR), cingulum angular bundle (CAB), cingulate gyrus endings (CCG), cortico-spinal tract (CST), inferior longitudinal fasciculus (ILF), superior longitudinal fasciculus parietal (SLFP) and temporal (SLFT) endings, and uncinate fasciculus (UNC).

The mean and standard deviation for each tract was computed for the CamCAN cohort. These normative values were then used to compute z-scores and corresponding p-values for each tract for each individual in both cohorts. The p-values for the 18 tracts for each subject were corrected for multiple comparisons [9] and then organized and plotted in Figure 2 to highlight the individual subjects with outlier activity values. The large preponderance of values less than the mean for the TBI cohort is evident in the right panel. The repeat reliability of the measures is apparent for those who had follow-up studies. Two or more tracts with $p < 10^{-3}$ was selected as the criterion to identify an individual as abnormal. With 63 TBI subjects, each with 18 tracts, i.e. 63 x 18 = 1134 tracts, this reduces the chance of a false positive to $p < 0.0012$ . For the normal group, i.e. 588 x 18 = 10,584 tracts, the chance of a false positive is $p < 0.012$ . One of the 588 controls and seven of the TBI cohort meet the criterion for abnormal white matter function.

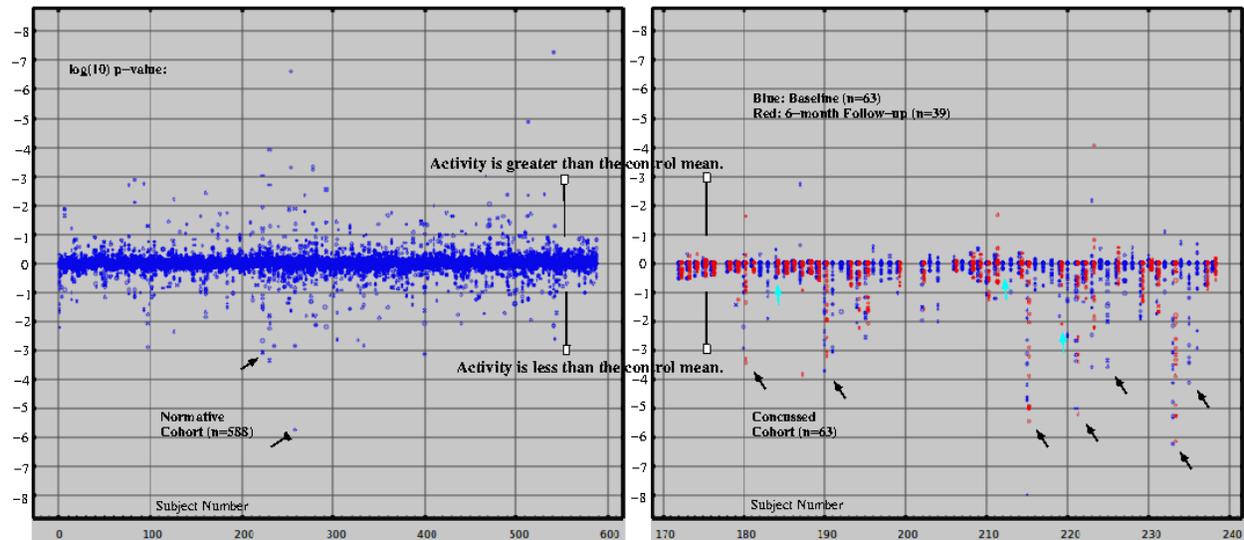

Figure 2. The activity within each of 18 tracts for each individual was converted to a z-score using the mean and standard deviation for each tract from the CamCAN (control) cohort. The p-values associated with the z-scores for each individual were corrected for multiple comparisons [9]. The logarithms$_{10}$ for these adjusted p-values are plotted for the 18 tracts for the members of the CamCAN cohort (left panel) and for the TBI cohort (right panel). The $\log_{10}$ p-value is plotted as a positive number for tracts whose neuroelectric activity was greater than the mean. Baseline and six-month follow-up (when available) are plotted side by side in blue and red respectively. Those who had at least two tracts for which $p < 10^{-3}$ were judged abnormal and are indicated with black arrows, $n = 7$ TBI, $n = 1$ controls and 1 control is close (blunt arrow). The cyan arrows indicate the three subjects who suffered severe TBI's.

**Conclusions and discussion**
(1) We demonstrate for the first time the capability to directly assess the functional integrity of white matter tracts for each individual including a probabilistic measure of confidence in the assessment for each tract.
(2) 60 of the 63 chronically symptomatic TBI cohort show no skull fractures, bleeding, or brain abnormalities on standard clinical brain imaging studies. 7 of the 60 did demonstrate a significant reduction in white matter neuroelectric activity (Figure 2). This result suggests a substantive increase in sensitivity to the abnormalities which likely underlie chronic symptoms of concussion. Without such sensitivity, the mechanism which underlies the majority of patients' symptoms and even the causal relationship to the triggering event, the concussion, remain completely unknown.
(3) The TBI cohort as a whole shows a significant decrease in activity in 10 of 18 white matter tracts (Figure 1). This is particularly noteworthy for the left and right uncinate fasciculi. It is implausible that this is due to undetected anatomic abnormalities in all or even most of the individuals. We therefore suggest an alternative to physical trauma as a cause of the systemic reductions in intercommunication between brain regions reported here, viz. the multiple supra-maximal stimuli coincident with the concussion. Almost all such stimuli converge on brainstem nuclei. That momentary multi-channel high-intensity sensory input may produce an avalanche of neuroelectric activity in the brainstem which produces persistent alterations in activity widely throughout the brain.

The measures reported by our group for cortical regions [7] and for white matter here are highly sensitive to such putative alterations in neuroelectric activity.
- (4) Trans-cranial magnetic stimulation (TMS) is a non-invasive and safe treatment modality [10] which likely alters persistent regional and systemic patterns of neuroelectric activity. We contemplate the use of MEG-derived measures of regional neuroelectric activity to provide individualized targeting, objective monitoring, and enhanced efficacy [11,12,13,14] of TMS treatment for symptoms of concussion, major depression and others.

**Acknowledgements**

We gratefully acknowledge the invaluable contributions made to this effort by the US Department of defense, the Cambridge (UK) Centre for Ageing and Neuroscience, the Extreme Science and Engineering Development Environment (XSede), the Open Science Grid (OSG), the San Diego Supercomputing Center, the Pittsburgh Supercomputing Center, the XSede Neuroscience Gateway, Darren Price, Mats Rynge, Rob Gardner, Frank Wurthwein, Derek Simmel, Mahidhar Tatineni, Ali Marie Shields.
Data used in the preparation of this work were obtained from the CamCAN repository [1,2]. The OSG is supported by the National Science Foundation, 1148698, and the US Department of Energy's Office of Science.